\documentclass{article}
\usepackage{IEEEtrantools}
\usepackage{spconf,amsmath,graphicx,hyperref}
\usepackage{orcidlink}
\usepackage{etoolbox}
\hypersetup{
  urlcolor     = blue, %Colour for external hyperlinks
  linkcolor    = blue, %Colour of internal links
}
\usepackage{cite}
\usepackage{booktabs}
\usepackage{graphicx}
\usepackage{newtxtext,newtxmath}
\usepackage{xspace}
\usepackage{xcolor}
\usepackage{url}
\usepackage{enumitem}

\newcommand{\cds}{\textsc{CDS}\xspace}
\newcommand{\hsc}{\textsc{HSC}\xspace}
\newcommand{\lsc}{\textsc{LSC}\xspace}
\newcommand{\msc}{\textsc{MSC}\xspace}
\usepackage{tikz}

\makeatletter
\patchcmd{\thebibliography}
  {\advance\leftmargin\labelsep}
  {\setlength{\itemsep}{1pt plus 0.2pt}
   \setlength{\parsep}{1pt}
   \setlength{\parskip}{1pt}
   \advance\leftmargin\labelsep}
  {}
  {}
\title{Distributional Metrics for Evaluating Spoken Conversational Systems}
\name{\parbox{\dimexpr\textwidth-2\tabcolsep\relax}{\centering
Shree Harsha Bokkahalli Satish\,\orcidlink{0009-0000-0554-7265}\textsuperscript{1},
Erica Cooper\,\orcidlink{0000-0002-2978-2793}\textsuperscript{2},
Patr\'{i}cia Schmidtov\'{a}\,\orcidlink{0009-0008-5516-798X}\textsuperscript{3}\\
Maike Z\"{u}fle\,\orcidlink{0009-0001-7238-7705}\textsuperscript{4},
\'{E}va Sz\'{e}kely\,\orcidlink{0000-0003-1175-840X}\textsuperscript{1},
Nicholas Sanders\,\orcidlink{0009-0007-0925-1606}\textsuperscript{5},
Ond\v{r}ej Klejch\,\orcidlink{0000-0001-5495-967X}\textsuperscript{5}
}}

\address{\parbox{\dimexpr\textwidth-2\tabcolsep\relax}{\centering
\textsuperscript{1}KTH Royal Institute of Technology, Sweden \textsuperscript{2}NICT, Japan \textsuperscript{3}Charles University, Czech Republic \textsuperscript{4}Karlsruhe Institute of Technology, Germany \textsuperscript{5}University of Edinburgh, UK\\
\href{mailto:shbs@kth.se}{\texttt{shbs@kth.se}}
}}
\begin{document}
\bstctlcite{no-author-dashes}
%\ninept
%
\maketitle
%

%\copyrightnoti
%\vspace{-0.5cm}    
\begin{abstract}
Evaluating conversational systems is a difficult and unresolved problem. We introduce the Conversational Distribution Score (\cds), which compares distributions of conversational behaviour using human conversations as a reference. \cds describes speech rate, syllabic rhythm, and turn interaction through eight interpretable features plus a separate two-feature semantic baseline. We compare conversations with two reference scales: one based on conversational success within human dialogue and another contrasting human and synthetic dialogue. Using listener judgments from out-of-domain goal--oriented dialogues, we examine system ranking, preferences between conversations, and ranking stability. Composite \cds recovers five of six listener system comparisons while individual features show strong correlation with listener preferences between conversations. We examine how many minutes and conversations are required before rankings stabilize. These findings support distributional comparisons as a complement to specific interactional metrics to evaluate conversations and conversational models while showing their interpretable value.
\end{abstract}

\begin{keywords}
conversation evaluation, full-duplex speech, distributional metrics, conversational AI
\end{keywords}

\section{Introduction and Background}
The rapid progress in speech synthesis~\cite{chen2025neural,le2023voicebox,tan2021survey} and conversational AI models~\cite{roy2026personaplex,openai2024realtime,google_gemini_live_api} has resulted in synthesis of isolated speech segments that sound natural and indistinguishable from real human speech~\cite{tan2024naturalspeech}. However, a conversation is not experienced as a set of isolated utterances~\cite{couper1996towards}.%A conversation's character arises from the distribution of not only the nature of these individual utterances but also the dynamics within such as speaking rate, turn length, response timing gaps, overlaps, prosody, affect and the way these behaviours change during the course of a conversation.

Recent work on evaluating conversational models include several benchmark suites and metrics.  The Full-Duplex-Bench series tests pause handling, backchanneling, turn--taking, user interruption using pre-recorded stimuli~\cite{lin2025fdb}, speech overlap~\cite{lin2026fdb15}, and multi--turn interaction with an automated examiner at different speaking rates~\cite{lin2026fdb2}. %, and multi-step tool use with disfluent speech~\cite{lin2026fdb3}.
MTR--DuplexBench~\cite{he2026mtr} and M3--DuplexBench~\cite{fukuda2026m3} extend this to longer conversations, additional dialogue dimensions, languages, and domains. SPEARBench evaluates broader naturalness dimensions through pre--recorded question--answer dialogues~\cite{thebaud2026spearbench}. %TurnNat assesses turn--taking naturalness through predicted two--speaker voice activity and evaluates its scores using human--validated timing perturbations~\cite{zhang2026turnnat}. 
These benchmarks provide us with controlled tests of specific interactional capabilities and task completion. 

However, our approach starts from the observation that conversational behaviour is naturally distributional and offers a complementary view to the above specific interactional evaluations. Because there are many ways to have realistic (and unrealistic) conversations, it becomes necessary to examine the full distribution of conversational behaviours. So, we ask a different evaluation question: \emph{If we collect conversations with a speech system, what do the distributions of its conversational behaviour tell us about that system in relation to the distribution of human conversations?}

Distributional metrics have been examined previously in Moshi~\cite{defossez2024moshi} and F-Actor~\cite{zufle-etal-2026-f}. Hallur \textit{et al.}~\cite{hallur2026reference} compare conversational prosody and rhythm with matched human reference groups. Additionally, in dGSLM, distributions of inter--pausal units, gaps, and overlaps in generated dialogue continuations were compared with those from Fisher conversations~\cite{nguyen2023generative}. However, the resulting statistics are not placed on a common reference scale, making it difficult to combine different behaviours or use them to compare systems. TTSDS~\cite{minixhofer2024ttsds} and TTSDS2~\cite{minixhofer2026ttsds2} address a related problem in synthetic speech generation by comparing distributions of synthetic and human reference speech. \cds brings these ideas into spoken interaction. We compare distances of several interpretable distributions to human references and answer the following research questions (RQs):
\begin{enumerate}[label=\textbf{RQ\arabic*:}, leftmargin=*, itemsep=3pt]
    \item \textit{How can distributional metrics distinguish spoken conversational systems?}
    \\(1) Do those same metrics also distinguish between human--human, and human--AI conversations?
    \\(2) Within human--AI conversations, do they distinguish between different conversational AI systems, and do the resulting system rankings agree with third-party human judgments?
    \item \textit{Which conversational features align most closely with human judgments?}
    \item \textit{How many callers and/or minutes of conversation are needed before metrics stabilize?}
\end{enumerate}
% \begin{enumerate}
% \item We introduce \cds as an interpretable way to compare conversational AI behaviour with human conversations and how they correlate with listener humanness ratings. 

% \item We distinguish conversational AI systems and measure how many callers and minutes are needed before listener rankings become stable.
% \end{enumerate}

\begin{figure}[!t]
\centering
\includegraphics[width=0.49\textwidth]{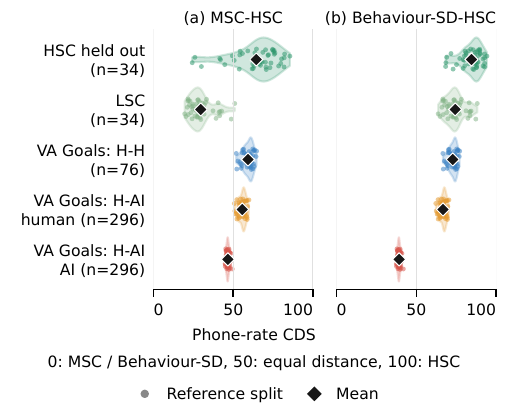}
\vspace{-0.8cm}
\caption{Phone-rate CDS scores for different conversation types. \cds scores indicate distributional similarity to the anchors.}
\label{fig:rq1}
\end{figure}

%%% ADdd urations of conversations above later %%%%%%%

\begin{figure*}[!t]
\centering
\includegraphics[width=0.98\textwidth]{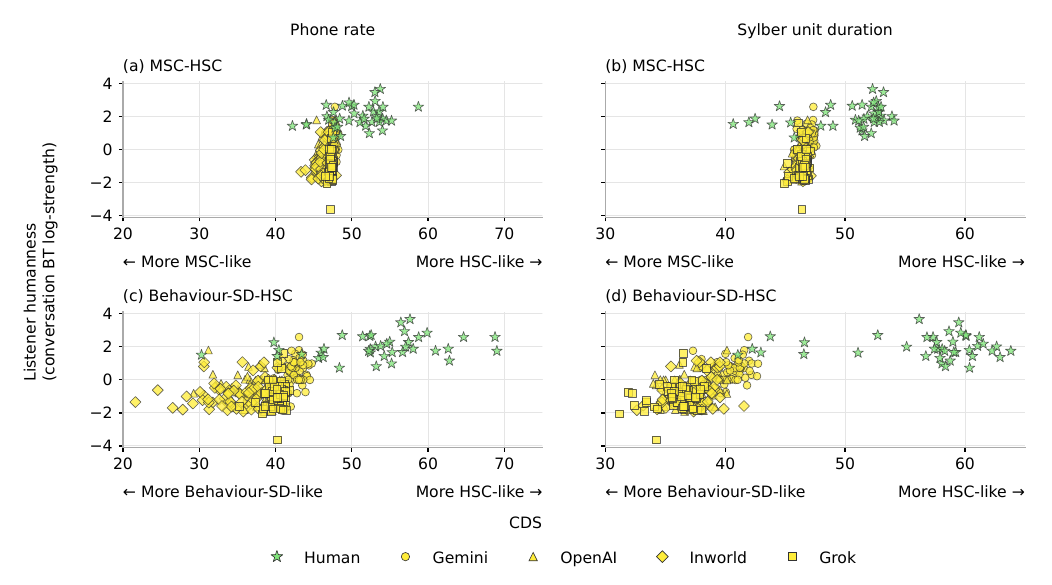}
\vspace{-0.6cm}
\caption{Listener ratings of humanness vs. phone rate and Sylber unit duration \cds for AI-partner and human-partner pooled turns per conversation. Each point is one conversation. AI scores cluster near MSC--HSC midpoint, while Behavior-SD--HSC moves AI scores closer to the synthetic distraction anchor.}
\label{fig:rq2}
\end{figure*}

\section{Conversational Distribution Score}\label{sec:cds}

Our goal is to score a conversational system using a set of conversations it has produced.
%Each conversation consists of a sequence of turns.
To evaluate a particular conversational behaviour, we can extract a feature $f$ that quantifies this behaviour from either each turn or even more granular levels and pool the features from all turns to form a distribution $T$. We define distributions for each conversation and we simply average these distributions to obtain a distribution for a system. We can then compare it with a reference distribution $R$ using the Wasserstein--2 distance. Since all our features $f$ are scalars, we compute it as: 

% For any group $G$ of conversations $c$, we extract eligible features. Then, we are able to compute the distribution of a feature $f$ at the conversation level by pooling all the turns $t$ (defined in Section~\ref{sec:method}) of that conversation or at the group level by pooling features from all turns of all conversations from a system together. Since all our features $f$ are scalars, we compute the exact weighted empirical Wasserstein--2 distance between two distributions $T$ and $R$:
\begin{equation}
d(T,R)=\left[\int_0^1
\left\{F^{-1}_{T}(u)-F^{-1}_{R}(u)\right\}^2du\right]^{1/2}
\label{eq:w2}
\end{equation}
%Here $F_{T,f}(z)=\sum_j w_j\mathbf{1}(x_j\leq z)$, 
where $F^{-1}$ is an inverse cumulative distribution function. 

Following TTSDS2, we are also interested in how the distributions stand in relation to two reference anchors. However, unlike TTSDS2, we allow both anchors to represent conversations, so that we can compare different conversational behaviours rather than speech against noise alone. For any distribution $T$, human reference $R$, and distractor distribution $A$ our \textbf{\cds} is therefore:
\begin{equation}
\operatorname{CDS}(T;A;R)=100 \times
\frac{d(T,A)}{d(T,R)+d(T,A)}.
\label{eq:cds}
\end{equation}
\noindent Hence 100 is reference--like, 0 is distractor--like, and 50 means that the target is equidistant from both $R$ and $A$.

%\vspace{-0.3cm}
\subsection{\cds Features}\label{sec:feats}
%We look at 10 features that broadly form four families:
We choose 10 features across four families that capture different aspects of conversational behaviour. These aspects need not vary together and cover speech pace, rhythm, turn-taking, and the semantic relationships and predictability of replies.
\begin{itemize}
  \setlength\itemsep{0.2em}
\item \emph{\textbf{Pace}} contains Allosaurus--recognized phones per second~\cite{li2020universal}, Sylber 0.1.4~\cite{cho2025sylber} segment rate, and the distribution of Sylber segment durations which can be thought as durations of syllable--like segments.

\item \emph{\textbf{Rhythm}} contains two features obtained from ordered Sylber durations $d_1,\ldots,d_n$ of the turn segments:
\begin{align}
\operatorname{nPVI}(d)&=\frac{100}{n-1}\sum_{i=1}^{n-1}
 \frac{|d_{i+1}-d_i|}{(d_{i+1}+d_i)/2}, \label{eq:npvi}\\
L_{\mathrm{final}}&=\log\frac{d_n}{\operatorname{median}(d_1,\ldots,d_{n-1})}.
\label{eq:lengthening}
\end{align}

%Positive $L_{\mathrm{final}}$ indicates a longer final syllable--like unit, which may signal turn yielding. while nPVI measures adjacent syllable duration variations adapted from music...\cite

% We use the original Sylber 0.1.4 model~\cite{cho2025sylber}, and its SSL segmentation provides a bridge from learned representations to interpretable features. 

\item \emph{\textbf{Turn Interaction}} contains turn duration, signed response gap between two speakers (silence is positive, while overlap is negative), and each speaker's share of combined conversation floor time.

\item \emph{\textbf{Lexical Content}} As a separate semantic baseline, we average \cds~(semantic) for whether the reply entails the preceding turns and its probabilities~\cite{poliak2020survey} from a RoBERTa model fine-tuned on InferConvAI~\cite{dziri2019evaluating} and GPT--2 mean response surprisal ($\log_2$ perplexity)~\cite{radford2019language} also using up to two preceding turns as context.
\end{itemize}

% \subsection{\cds Weighting}\label{sec:pooling}
% For each feature, we average the distributions of the $N$ conversations with valid measurements as $P_G=\frac{1}{N}\sum_{c=1}^{N}P_c.\label{eq:pool}$
% Here $P_c$ is the weighted empirical distribution for conversation $c$. We weight speakers equally, then valid turns within speakers and additionally for Sylber duration, units within turns.
% Response gaps and floor shares receive equal weight across retained transitions and speakers, respectively. The quantile function used in Eq.~\ref{eq:w2} is the generalized inverse of the mixture cumulative distribution function (CDF)~\cite{embrechts2013note}.

\section{Datasets}

\textbf{Human references:} We use \texttt{CANDOR}, a corpus of $>=$25~min dyadic, dual-channel human conversations with post conversation surveys from both participants~\cite{reece2023candor}. Following the perceived conversation success analysis of Withanage \textit{et al.}~\cite{withanage2026acoustic}, we form high-success (91) and low-success (35) (\hsc and \lsc) groups based on the post--conversation survey measures filled by the participants. After removing cross-label speakers, our analysis retained 88 \hsc conversations and 34 \lsc conversations. We also separately froze 35 participant--disjoint medium success (\msc) conversations as a middle ground. 

\noindent \textbf{VoiceArena Goals \texttt{(VA Goals)} Dataset:} VoiceArena~\cite{voicearena2026} provided us with 374 conversations involving 56 human callers ranging from $\approx$ 2 to 13~mins, five dyadic interaction scenarios (\textit{Booking a flight for two people}, \textit{Rescheduling a passenger on a flight}, \textit{Preparing for a friend's wedding}, \textit{Dogs vs. cats debate}, \textit{Food preferences clarification} (100, 100, 72, 66, 36 conversations respectively)) and five speaking partners: four spoken conversational AI systems (Gemini~\cite{google_gemini_live_api}, OpenAI~\cite{openai2024realtime}, Inworld~\cite{inworld_realtime_api}, and Grok~\cite{xai2025grokvoice}) and another human partner as control. This left us with 296 human--AI and 76 human--human conversations. Separate third-party listeners also provided us with 1,026 \emph{humanness} pair--wise judgment votes for the four AI systems over 295 unique conversation pairs. We also used 120 interactions from Behavior--SD~\cite{lee2025behavior} set to serve as AI--AI distractor anchors to compare against.

\section{Methodology and Experiments}\label{sec:method}

We apply Silero Voice Activity Detection (VAD) 6.2.1~\cite{silero2026vad} to get turns from each speaker. We intentionally use simpler VAD methods to extract turns, over content related methods, for better generalization across languages and datasets. We use segments that have a 3~s minimum length and a 30~s maximum length for the pace and rhythm features. 700~ms of silence constitutes a new turn. Turn duration, response gaps, overlap, and floor time  use the complete speech timeline as is. % This distinctions prevents the acoustic duration filters from changing the turn--interaction features that we intend to measure.

We use 50 splits from \hsc for $R$ and use \msc and Behavior--SD as two choices of distraction anchors $A$. We expect \hsc--\msc to highlight differences in conversational behaviour associated with perceived conversational success. \hsc--Behavior--SD scales can reflect differences in how `natural' the speech sounds. We average scores across the 50 splits, across features within pace, rhythm, and interaction families in Section~\ref{sec:feats} and finally average the three non-semantic families equally to get the final \cds score.

We organize our experiments around: 

\textbf{RQ1}: (1) We compare human--human, human--AI feature distributions within each group and calculate \cds under both reference scales. (2) We then ask whether \cds and listeners rank the four systems in \texttt{VA Goals} similarly. Since systems were evaluated with different mixtures of callers and scenarios, we allow each caller/scenario/system block $b$ its own baseline. For a conversation score $z_{bp}$ from system $p$, we fit
\begin{equation}
z_{bp}=\alpha_b+\beta_p+\varepsilon_{bp},
\qquad \sum_p\beta_p=0.
\label{eq:adjustment}
\end{equation}
Here $\alpha_b$ is the block's baseline and $\beta_p$ is the system's adjusted contribution, which we use to rank systems. We derive the listener ranking from pairwise votes using a Bradley--Terry (BT) model. Each system receives a preference rate $\theta_p$, with higher values indicating a greater probability of being preferred:
$P(p\succ q)=\{1+\exp(\theta_q-\theta_p)\}^{-1}.
\label{eq:bt}$
Ties contribute half a win. We also report how many of the six system pairs favour the same system under \cds and listener strengths.

\textbf{RQ2}: We compare individual features within \cds using both system--ranking agreement and correlations with human preferences. For each matched conversation pair $(A,B)$, let $p_A$ be the share of votes favouring A, counting ties as half a vote. Let $C$ denote a feature or composite score averaged across the \hsc splits. We calculate Spearman correlation between $C(A)-C(B)$ and $p_A-0.5$. A positive correlation means that a larger \cds advantage tends to accompany a stronger listener preference for the same conversation. We repeat these analyses with both reference scales and with TTSDS2.

\begin{figure*}[t]
\centering
\includegraphics[width=0.925\textwidth]{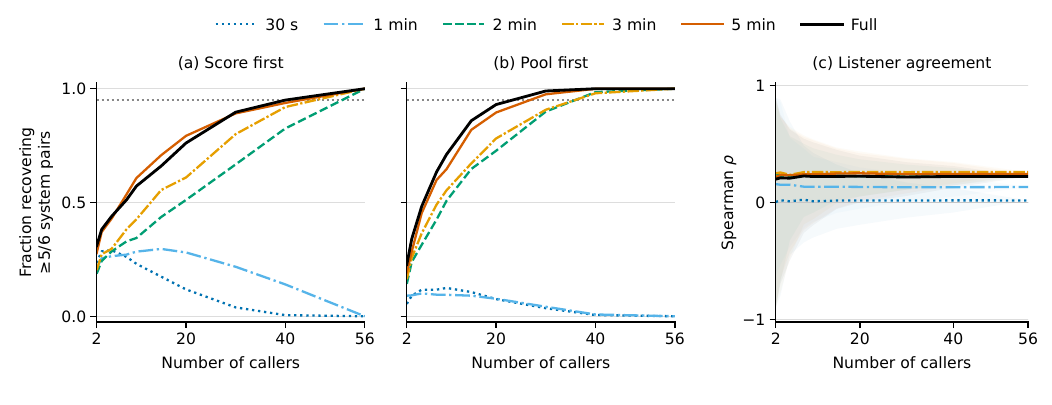}
\vspace{-0.8cm}
\caption{Effect of caller count and recording duration under HSC–MSC. Subfigures (a,b) show the fraction of subsets recovering at least five of six system comparisons in each estimator’s full-recording, all-caller ordering. Subfigure (c) shows the median correlation between overall conversation CDS differences and listener preferences.}
\label{fig:rq3}
\end{figure*}

\textbf{RQ3}: We sample callers from the \texttt{VA Goals} dataset 2000 times without replacement and evaluate the first 30~s to the full conversation and keep track what percentage retain at least 5/6 pair orderings found from full \cds:\hsc-\msc. In \emph{score first}, we calculate \cds for each conversation, average within each caller/scenario/system block, then over all scenarios and finally over callers for each system. In \emph{pool first}, we combine each system's turns into one distribution before calculating \cds, giving each conversation same total weight.

\begin{table}[!b]
\centering
\caption{Agreement with listener preferences and rankings. Spearman’s \(\rho\) relates score differences to average listener preferences for conversation pairs.%Ranges show the middle 95\% of correlations after resampling.
}
\label{tab:normalization}
% Pooled estimates
\setlength{\tabcolsep}{3pt}
\setlength{\tabcolsep}{2pt}
\begin{tabular*}{\linewidth}{@{}l@{\extracolsep{\fill}}rrc@{}}
\hline
Measurement & $\rho$ & \shortstack{95\% caller range} & \shortstack{Pairs correct} \\
\hline
\multicolumn{4}{@{}l}{\emph{\cds: HSC--MSC}} \\
Overall (non-semantic) & $0.222$ & $[0.097, 0.353]$ & 5/6 \\
Sylber duration & $0.503$ & $[0.412, 0.587]$ & 4/6 \\
Semantic & $0.083$ & $[-0.008, 0.182]$ & 4/6 \\
\hline
\multicolumn{4}{@{}l}{\emph{\cds: HSC--Behavior-SD}} \\
Overall (non-semantic) & $0.266$ & $[0.166, 0.368]$ & 3/6 \\
Sylber duration & $0.615$ & $[0.538, 0.681]$ & 5/6 \\
Semantic & $-0.240$ & $[-0.340, -0.128]$ & 1/6 \\
\hline
\multicolumn{4}{@{}l}{\emph{TTSDS2: HSC vs noise}} \\
Overall & $-0.062$ & $[-0.153, 0.032]$ & 4/6 \\
Wav2Vec2 activations & $0.555$ & $[0.482, 0.625]$ & 5/6 \\
D-vector & $0.419$ & $[0.317, 0.521]$ & 6/6 \\
\hline
\end{tabular*}
\end{table}
\section{Results and Discussion}
\textbf{(RQ1):} Under HSC--MSC, pooled phone-rate \cds averages 59.2 for human--human conversations, 46.6 for AI partners (Fig.~\ref{fig:rq1}). With Behavior--SD, these scores become 72.9 and 39.2. A \cds score $>$ 50 indicates that the distribution of that group is more \hsc--like than the distraction anchor. 

We also fit conversation-level BT models within each connected caller/scenario comparison group, including human partners. Fig.~\ref{fig:rq2} shows that many AI phone rate and Sylber duration distributions are nearly equidistant on HSC--MSC scale whereas a HSC--Behavior--SD scale shifts and spreads the AI scores more towards Behavior--SD indicating higher similarity to the distraction anchor. 

\noindent \textbf{(RQ2):} Individual \cds features identify interpretable features whose distributions can be inspected. From Table~\ref{tab:normalization}, we find that Sylber duration has higher preference correlation than overall \cds under both anchors. Similarly, Wav2Vec2 activations reach $\rho=0.555$ while overall TTSDS2 has $\rho=-0.062$. Wav2Vec2 activations provide a useful evaluation signal, but these scores do not directly identify which behaviour differs from references. Although Sylber duration \cds correlates strongly with listener preferences, which might indicate what current AI systems lack, we don't claim that matching duration distributions alone would improve perceived humanness. %Reporting our pace, rhythm, and interaction features allows us to examine conversational behaviour beyond a single number.
% Whether changes to these features improve perceived humanness we leave for further wrok.

\noindent \textbf{(RQ3)}: With full recordings, when we pool from all conversations, we can recover 99.0\% at 30 callers (1980/2000 subsets). Whereas if we score each conversation first and then the average reaches 95.0\% at 40 (1,900/2,000). Both methods are unable to recover five of six pairwise comparisons with all 56 callers using 30 s or 1 min (Fig.~\ref{fig:rq3}). Additional results for other features, code and datasets will be released on this \href{https://shreeharsha-bs.github.io/conversational-distribution-score}{website}.
% \begin{table*}
% \setlength{\tabcolsep}{5pt}
% \renewcommand{\arraystretch}{1.15}
% \begin{tabular*}{\linewidth}{@{}l@{\extracolsep{\fill}}rrrr@{}}
% \hline
% \multicolumn{5}{@{}l}{\textit{System-pooled overall \cds: mean $\pm$ SD across 50 HSC splits}} \\
% \hline
%  & \multicolumn{2}{c}{HSC--MSC} & \multicolumn{2}{c@{}}{HSC--Behaviour-SD} \\
% \cline{2-3}\cline{4-5}
% System ($n$) & \shortstack{Hierarchical\\weights} & \shortstack{Equal observation\\weights} & \shortstack{Hierarchical\\weights} & \shortstack{Equal observation\\weights} \\
% \hline
% Gemini (73) & $48.52 \pm 1.97$ & $48.26 \pm 1.82$ & $44.43 \pm 1.37$ & $43.96 \pm 1.33$ \\
% OpenAI (76) & $48.68 \pm 1.98$ & $48.27 \pm 1.84$ & $44.38 \pm 1.34$ & $43.11 \pm 1.32$ \\
% Inworld (75) & $48.13 \pm 1.98$ & $47.25 \pm 1.89$ & $45.01 \pm 1.52$ & $42.90 \pm 1.57$ \\
% Grok (72) & $47.76 \pm 1.95$ & $47.71 \pm 1.79$ & $43.93 \pm 1.64$ & $43.56 \pm 1.55$ \\
% \hline
% Listener system pairs agreement & 5/6 & 4/6 & 4/6 & 4/6 \\
% \hline
% \end{tabular*}

% \end{table*}

\section{Conclusion}
We introduced \cds to compare distributions of conversational behaviour using human conversations as references. Its individual features describe specific behavioural differences, while the composite summarizes the selected feature families.
On \texttt{VA Goals} dataset, agreement with listener judgments depends on the reference scale and whether we evaluate system ordering or preferences between conversations. Short excerpts of one minute or less fail to reproduce rankings reliably. These findings support \cds as an interpretable complement to evaluating conversational models. Future work will involve looking at more features, examining other conversational behaviours and deriving a more nuanced composite score.

\clearpage
\section{Acknowledgments}
This work was supported by JSALT 2026 at JHU with funds from NSF CCRI Grant No. 2120435, Google DeepMind, JHU HLTCOE, JHU AI2AI and ACL, by the Wallenberg AI, Autonomous Systems and Software Program (WASP) funded by the Knut and Alice Wallenberg Foundation, and by GAUK 252986 and SVV 260 698 and by the EU (ERC, NG-NLG, 101039303). We used OpenAI Codex to assist with statistical-analysis code, and generating plots from data which were then reviewd and verified by the authors.

\vspace{-0.25cm}
\bibliographystyle{IEEEtran}
\bibliography{strings,refs}

\end{document}